\documentclass{ifacconf}

\usepackage{graphicx}      % include this line if your document contains figures
\usepackage{natbib}        % required for bibliography
\usepackage{amsmath}
\usepackage{amssymb}
\usepackage{url}

\newtheorem{assumption}{Assumption}

\begin{document}
\begin{frontmatter}

\title{From Nonlinear Identification to Linear Parameter Varying Models: Benchmark Examples\thanksref{footnoteinfo}} 
% Title, preferably not more than 10 words.

\thanks[footnoteinfo]{This work has received funding from the European Research Council (ERC) under the European Union's Horizon 2020 research and innovation programme (grant agreement nr. 714663).}

\author[First]{Maarten Schoukens} 
\author[First]{Roland T\'oth}

\address[First]{Control Systems, Eindhoven University of Technology, Eindhoven, The Netherlands (e-mail: m.schoukens@tue.nl, r.toth@tue.nl).}

\begin{abstract}                % Abstract of not more than 250 words.
	Linear parameter-varying (LPV) models form a powerful model class to analyze and control a (nonlinear) system of interest. Identifying a LPV model of a nonlinear system can be challenging due to the difficulty of selecting the scheduling variable(s) a priori, which is quite challenging in case a first principles based understanding of the system is unavailable.
	
	This paper presents a systematic LPV embedding approach starting from nonlinear fractional representation models. A nonlinear system is identified first using a nonlinear block-oriented linear fractional representation (LFR) model. This nonlinear LFR model class is embedded into the LPV model class by factorization of the static nonlinear block present in the model. As a result of the factorization a LPV-LFR or a LPV state-space model with an affine dependency results. This approach facilitates the selection of the scheduling variable from a data-driven perspective. Furthermore the estimation is not affected by measurement noise on the scheduling variables, which is often left untreated by LPV model identification methods.
		
	The proposed approach is illustrated on two well-established nonlinear modeling benchmark examples.
\end{abstract}


\end{frontmatter}
%===============================================================================

\section{Introduction}
	The Linear-Parameter Varying (LPV) framework offers a powerful tool set to model and control nonlinear systems \citep{Mohammadpour2012}. The need to identify high-quality LPV models is high due to their use in many control applications. The identification of LPV models has been studied in detail \citep{Toth2010,Santos2011}. However, most LPV identification approaches assume the knowledge of the scheduling signal a priori and/or have to rely on noisy measurements of this variable.
	
	Embedding nonlinear models into the LPV model class offers an alternative approach to obtain LPV models of nonlinear systems without having to identify a LPV model directly. This avoids the selection of appropriate scheduling signals during the LPV identification: the scheduling signal(s) are obtained as a result of the nonlinear embedding. Although the embedding of nonlinear systems into the LPV framework is a popular approach in the control of nonlinear systems, only a few systematic embedding methods are discussed in the literature \citep{Chisci2003,Young2011,Mohammadpour2012,Abbas2014,Abbas2017}. Such a systematic embedding approach allows to widen the class of nonlinear systems that can be effectively controlled using the LPV framework.
	
\begin{figure}[bt]
		\centering
			\includegraphics[width=0.95\columnwidth]{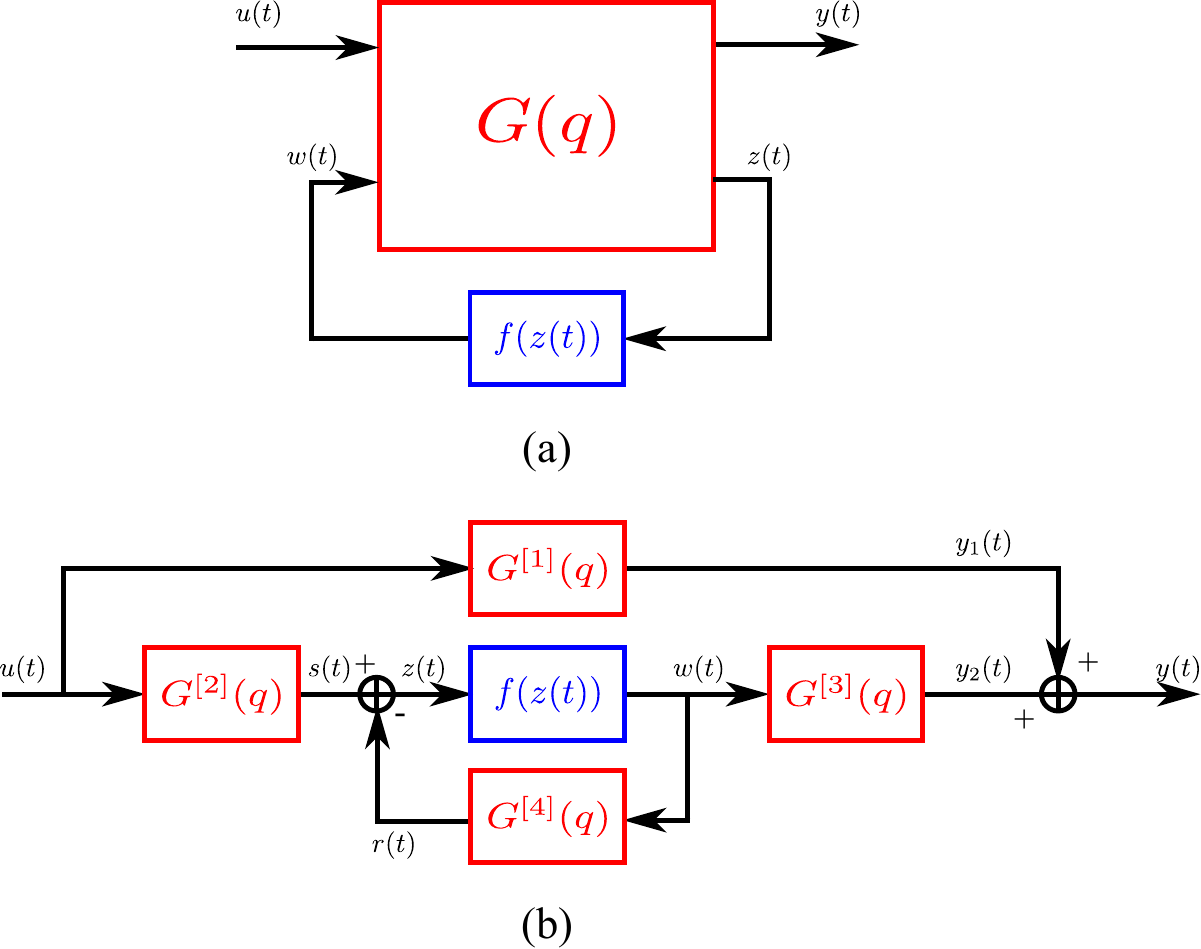}
		\caption{The nonlinear LFR structure represented by (a) a two-input two-output LTI block $G(q)$, and (b) an equivalent block-oriented structure with 4 SISO LTI blocks $G^{[i]}(q)$.}
		\label{fig:LFR}
	\end{figure}

	This paper presents a systematic nonlinear system embedding approach for nonlinear systems represented by LFRs with a static nonlinear block (see Figure~\ref{fig:LFR}) \citep{SchoukensM2017b}. It is shown how one can first identify a nonlinear LFR model, and embed this model in an automated and systematic way into a LPV representation such that a LPV-LFR, or alternatively an affine state-space LPV model results, without introducing new singularity points in the representation.
	
	In the following sections, the nonlinear LFR system class is discussed first (Section~\ref{sec:LFRClass}). The embedding of these nonlinear systems into a LPV representation is discussed next in Section~\ref{sec:Embedding}. Finally two benchmark examples are given to illustrate how one can go from nonlinear identification to a LPV system representation in a systematic way (Sections~\ref{sec:WH} and~\ref{sec:Silverbox}). Finally, some conclusions are presented in Section~\ref{sec:Conclusions}.
	
\section{Nonlinear LFR System Class} \label{sec:LFRClass}
	The considered class of systems is the class of nonlinear systems that can be represented by a single-input single-output (SISO) discrete-time nonlinear linear fractional representation (LFR). The nonlinear LFR system class is a general system class comprising only one localized SISO static nonlinearity (see Figure~\ref{fig:LFR}) \citep{Vandersteen1999,Vanbeylen2013}. Generality is based upon the fact that widely used block-oriented Hammerstein, Wiener and Wiener-Hammerstein structures are special cases of the nonlinear LFR structure (see \citep{Giri2010,SchoukensM2017b} for more information on the identification of block-oriented structures and some of their applications). The SISO case is considered here for simplicity. The presented results can also extended to the MIMO setting.
		
	The input-output relation of a nonlinear LFR, where each of the subsystems represented by $G^{[i]}$ is assumed to be a discrete-time and causal LTI system, can be expressed in (minimal) state-space form as:
	\begin{align}
		\begin{split} \label{eq:LFRSS1}
		x(t+1) &= A x(t) + B_\mathrm{u} u(t) + B_\mathrm{w} w(t), \\
		z(t) &= C_\mathrm{z} x(t) + D_\mathrm{yu} u(t) + D_\mathrm{yw} w(t), \\
		y(t) &= C_\mathrm{y} x(t) + D_\mathrm{zu} u(t) + D_\mathrm{zw} w(t), \\
		w(t) &= f(z(t)),
		\end{split}
	\end{align}
	where $f(z(t))$ is a SISO static nonlinear function, and $A$, $B_\mathrm{u}$, $B_\mathrm{w}$ $C_\mathrm{z}$, $C_\mathrm{y}$, $D_\mathrm{yu}$, $D_\mathrm{yw}$, $D_\mathrm{zu}$, $D_\mathrm{zw}$ are real matrices of appropriate dimensions. Eliminating $z(t)$ from these equations results in:
	\begin{align}
		\begin{split} \label{eq:LFRSS2}
		&x(t+1) \!=\! A x(t) \!+\! B_\mathrm{u} u(t) \!+\! B_\mathrm{w} f(C_\mathrm{z} x(t) \!+\! D_\mathrm{zu} u(t) \!+\! D_\mathrm{zw} w(t)), \\
		&y(t)   \!=\! C_\mathrm{y} x(t) \!+\! D_\mathrm{yu} u(t) \!+\! D_\mathrm{yw} f(C_\mathrm{z} x(t) \!+\! D_\mathrm{zu} u(t) \!+\! D_\mathrm{zw} w(t)), \\
		&w(t) \!=\! f(C_\mathrm{z} x(t) \!+\! D_\mathrm{zu} u(t) \!+\! D_\mathrm{zw} w(t))
		\end{split}
	\end{align}
	To also eliminate $w(t)$, it is assumed that at least a one-sample delay is present in the feedback dynamics $G^{[4]}(q)$ (in other words, $D_\mathrm{zw}=0$), where $q^{-1}$ is the backwards shift operator, see Figure~\ref{fig:LFR}. This results in the following simplified expression:
	\begin{align}
		\begin{split} \label{eq:LFRSS3}
		x(t+1) &= A x(t) + B_\mathrm{u} u(t) + B_\mathrm{w} f(C_\mathrm{z} x(t) + D_\mathrm{zu} u(t)), \\
		y(t)   &= C_\mathrm{y} x(t) + D_\mathrm{yu} u(t) + D_\mathrm{yw} f(C_\mathrm{z} x(t) + D_\mathrm{zu} u(t)),
		\end{split}
	\end{align}

\section{LPV Embedding} \label{sec:Embedding}
	This section illustrates how nonlinear system representations belonging the nonlinear LFR system class can be transformed into a LPV affine state-space representation.

	The static nonlinearity that is present in the nonlinear model needs to be factorized to obtain a LPV representation. Multiple factorization approaches are possible. Here, the nonlinear function $f(z(t))$ is decomposed as $z(t)\bar{f}(z(t)) + c$.

	\begin{assumption} \label{ass:SNL} The static nonlinear function $f(z(t))$ can be represented as: $z(t)\bar{f}(z(t)) + c$, such that $\bar{f}(z(t))$ does not contain singular points in the region of interest, and $c$ is finite. $\bar{f}(z(t))$ is the scheduling map. \end{assumption}
	This assumption excludes functions $f(z(t))$ that have singularities in the region of interest, e.g. $\frac{1}{z}$ if $z=0$ lies within the range of interest.
	
	Under Assumption~\ref{ass:SNL}, the nonlinear LFR structure with $D_\mathrm{zw}=0$ can be represented by:
	\begin{align}
		\begin{split}
		x(t+1) &= A x(t) + B_\mathrm{u} \tilde{u}(t) + B_\mathrm{w} \tilde{f}(C_\mathrm{z} x(t) + D_\mathrm{zu} \tilde{u}(t)), \\
		\tilde{y}(t)   &= C_\mathrm{y} x(t) + D_{yu} \tilde{u}(t) + D_\mathrm{yw} \tilde{f}(C_\mathrm{z} x(t) + D_\mathrm{zu} \tilde{u}(t)),
		\end{split}
	\end{align}
	where:
	\begin{align} \label{eq:offsetLPV}
		\begin{split}
		\tilde{f}(z(t)) &= z(t) \bar{f}(z(t)) \\
		\tilde{u}(t) &= u(t) - \frac{G_0^{[4]}}{G_0^{[2]}} c  \\
		\tilde{y}(t) &= y(t) - \left( G_0^{[3]} + \frac{G_0^{[1]}G_0^{[4]}}{G_0^{[2]}} \right) c  \\
		G^{[1]}_0 &= D_\mathrm{yu} + C_\mathrm{y} (I-A)^{-1} B_\mathrm{u} \\
		G^{[2]}_0 &= D_\mathrm{zu} + C_\mathrm{z} (I-A)^{-1} B_\mathrm{u} \\
		G^{[3]}_0 &= D_\mathrm{yw} + C_\mathrm{y} (I-A)^{-1} B_\mathrm{w} \\
		G^{[4]}_0 &= D_\mathrm{zw} + C_\mathrm{z} (I-A)^{-1} B_\mathrm{w}
		\end{split}
	\end{align}
	Of course, for Eq.~\eqref{eq:offsetLPV} to hold $\frac{G_0^{[4]}}{G_0^{[2]}}$ and $\left( G_0^{[3]} + \frac{G_0^{[1]}G_0^{[4]}}{G_0^{[2]}} \right)$ should be finite.
	
	Under Assumption~\ref{ass:SNL} we can represent the nonlinear LFR structure as a LPV affine state-space representation with an additional constant offset at the input and output:
	\begin{align} \label{eq:LFRSS4}
		\begin{split}
		x(t+1) &= A x(t) + B_\mathrm{u} \tilde{u}(t) + B_\mathrm{w} p(t)\left(C_\mathrm{z} x(t) + D_\mathrm{zu} \tilde{u}(t)\right), \\
		\tilde{y}(t)   &= C_\mathrm{y} x(t) + D_\mathrm{yu} \tilde{u}(t) + D_\mathrm{yw} p(t)\left(C_\mathrm{z} x(t) + D_\mathrm{zu} \tilde{u}(t)\right),
		\end{split}
	\end{align} 
	where $p(t)$ is given by the scheduling map:
	\begin{align}
		\begin{split}
		p(t) &= \bar{f}(z(t)),
		\end{split}
	\end{align}
	with $z(t)$ satisfying Eq.~\eqref{eq:LFRSS1}.
	
	This results in the following affine state-space LPV structure:
	\begin{align} \label{eq:AffineLPV_LFR}
		\begin{split}
		x(t+1) &= A x(t) + A_\mathrm{p} p(t) x(t) + B_\mathrm{u} \tilde{u}(t) + B_\mathrm{p} p(t) \tilde{u}(t), \\
		\tilde{y}(t)   &= C_\mathrm{y} x(t) + C_\mathrm{p} p(t) x(t) + D_\mathrm{yu} \tilde{u}(t) + D_\mathrm{p} p(t) \tilde{u}(t),
		\end{split}
	\end{align}
	with:
	\begin{align}
		\begin{split}
		A_\mathrm{p} &= B_\mathrm{w} C_\mathrm{z},\\
		B_\mathrm{p} &= B_\mathrm{w} D_\mathrm{zu},\\
		C_\mathrm{p} &= D_\mathrm{yw} C_\mathrm{z},\\
		D_\mathrm{p} &= D_\mathrm{yw} D_\mathrm{zu}.
		\end{split}
	\end{align}
	
	\subsection{Remarks} \label{sec:remarks}
	
	From a control design point of view it is desirable for the scheduling signal $p(t)$ or the input of the static nonlinearity $z(t)$ to be measurable or observable. However, in the current work, $p(t)$ is obtained from the nonlinear model, by the static nonlinear function $\bar{f}$. In some cases a measurable signal $z(t) = \tilde{y}(t)$ is obtained, e.g. when $G^{[1]} = G^{[2]}$ and $G^{[3]} = -G^{[4]}$. In other cases, one can look for state transformations such that $z(t)$ depends only on measurable/observable states, the (past) input and output \citep{Abbas2017}.
	
	A similar approach as the one presented in this paper can be used to obtain a LPV representation of nonlinear structures containing multiple static nonlinear blocks, at the cost of introducing multiple scheduling variables.
	
	The constant offset at both the input and the output can be dealt with during the LPV control design process as a unity disturbance or by using input or output trimming methods.
	
	The affine state-space LPV representation is not unique: a state transformation can be introduced. Furthermore, the latent variables $z(t)$ and $w(t)$ are also not unique. This is similar to the non-uniqueness in the representation of block-oriented systems \citep{SchoukensM2015a,SchoukensM2015c}. Hence, there are degrees of freedom in the proposed method that can be further exploited especially in the MIMO setting.

\section{Wiener-Hammerstein Benchmark} \label{sec:WH}
	The Wiener-Hammerstein benchmark \citep{Schoukens2009a} is a well-established nonlinear system identification benchmark. It features a Wiener-Hammerstein block-oriented electronic circuit: a one-sided saturation diode-resistor nonlinearity is sandwiched in between two low-pass linear filters. Only noise at the output of the system is present (SNR$\approx$60dB). The benchmark data and more information on the benchmark system and previous benchmark results is available through: \url{www.nonlinearbenchmark.org}.
	
	Note that a Wiener-Hammerstein system class is a subset of the nonlinear LFR system class used in this paper where $G^{[1]}(q)$ and $G^{[4]}(q)$ are equal to zero.
	
	\subsection{Nonlinear Identification}
		A block-oriented Wiener-Hammerstein model is estimated using the approach presented in \citep{Sjoberg2012}. Many other identification approaches exist for this type of structures, an overview is given in \citep{SchoukensM2017b}.
		
		In a first step, a 6th order linear approximation of the nonlinear system is estimated. Secondly, the poles and zeros are allocated either to the front or the back LTI subsystem using a pole-zero allocation scan, a low-order polynomial model (up to power 3) of the nonlinearity $f(z)$ is estimated simultaneously. The low order polynomial nonlinearity is replaced by a neural network representation next (10 $\tanh$ neurons) to improve the model quality. Finally, all the model parameters are optimized together, minimizing the squared difference between the measured and the modeled output.
		
		Note that, due to the non-uniqueness of the input-output Wiener-Hammerstein representation (see e.g. \citep{SchoukensM2017b}) a scaled version of the nonlinearity $\hat{f}(z) = \alpha f(\beta z)$ could be obtained, compensated for by scaling the LTI dynamics: $\hat{G}^{[2]}(q) = \frac{1}{\beta} G^{[2]}(q)$, $\hat{G}^{[3]}(q) = \frac{1}{\alpha} G^{[3]}(q)$. Additionally, one could also substract a linear gain $\gamma$ from the nonlinearity $\hat{\hat{f}}(z) = f(z) -\gamma z$ and compensate for this by modifying the LTI dynamics $G^{[1]}(q) = \gamma G^{[2]}(q)G^{[3]}(q)$ without violating the LFR nature of the considered model class. However, the impact of these non-uniqueness transformations is not  explored further in this paper.
	
	\subsection{LPV Embedding} \label{sec:WHEmbedding}
		During the LPV Embedding, a constant $c = f(0)$ is extracted. Next, the function $f(z(t))-c$ is re-fitted with a model of the form $z(t)\bar{f}(z(t))$ where $\bar{f}(z(t))$ is again represented by a neural network (10 radial basis function neurons). Theoretically, such a re-fitting would not be required, one could simply define $\bar{f}(z) = \frac{f(z)-c}{z}$. Practically this could lead to undesired effects if $c = f(0)$ is determined with a limited accuracy, and the chosen nonlinearity description does not allow for a symbolic simplification of $\bar{f}(z)$ (as is for instance possible for polynomial expressions).
		
		The LPV embedding as described in Section~\ref{sec:Embedding} is performed next. The constant term $c$ is moved to the output of the LPV system $\bar{c} = cG^{[3]}_0$. The obtained LPV representation is given by:
		\begin{align}
			x(t+1) &\!=\! A x(t) \!+\! A_\mathrm{p} p(t) x(t) \!+\! B_\mathrm{u} u(t) \!+\! B_\mathrm{p} p(t) u(t),  \\
			y(t)   &\!=\! C_\mathrm{y} x(t) \!+\! C_\mathrm{p} p(t) x(t) \!+\! D_\mathrm{yu} u(t) \!+\!  D_\mathrm{p} p(t) u(t) \!+\! \bar{c}. \nonumber
		\end{align}
		
	\subsection{Results}
		The resulting LPV model parameters are given by (up to 4 digits):
		\begin{align}
			A &= \left[\begin{array}{cccccc} 
					2.1864  &-1.7852  &0.5254  &0        &0         &0 \\
					1       &0        &0       &0        &0         &0 \\
					0    	&1        &0       &0        &0         &0 \\
					0       &0        &0       &2.6005   &-1.1406   &0.6733 \\
					0       &0        &0       &2        &0         &0 \\
					0       &0        &0       &0    	 &0.5       &0 \end{array}\right], \nonumber\\
			A_\mathrm{p} &= \left[\begin{array}{cccccc} 
					0       &0        &0       &0        &0         &0 \\
					0    	&0        &0       &0        &0         &0 \\
					4.8398  &-1.2436  &0.6001  &0        &0         &0 \\
					0       &0        &0       &0        &0         &0 \\
					0       &0        &0       &0        &0         &0 \\
					0       &0        &0       &0    	 &0         &0 \end{array}\right], \quad B_\mathrm{u} = \left[\begin{array}{c} 2 \\ 0 \\0 \\0 \\ 0 \\ 0 \end{array}\right], \nonumber
		\end{align}
		\begin{align}			
			B_\mathrm{p} = \left[\begin{array}{c} 0 \\ 0 \\ 0 \\ 1.3314 \\ 0 \\ 0 \end{array}\right], \quad
			D_\mathrm{yu} = 0, \quad D_\mathrm{p} = -0.0364, \nonumber
		\end{align}
		\begin{align}
			C_\mathrm{y} &= \left[\begin{array}{cccccc}0 &0  &0 &-2.3773 &1.8094 &-2.3805 \end{array}\right], \nonumber\\
			C_\mathrm{p} &= \left[\begin{array}{cccccc}-0.1323& 0.0340& -0.0164& 0& 0&  0 \end{array}\right], \nonumber\\
			c &= -9.5254 \; 10^{-4}.
		\end{align}
		
\begin{figure}[hbt]
			\centering
				\includegraphics[width=0.95\columnwidth]{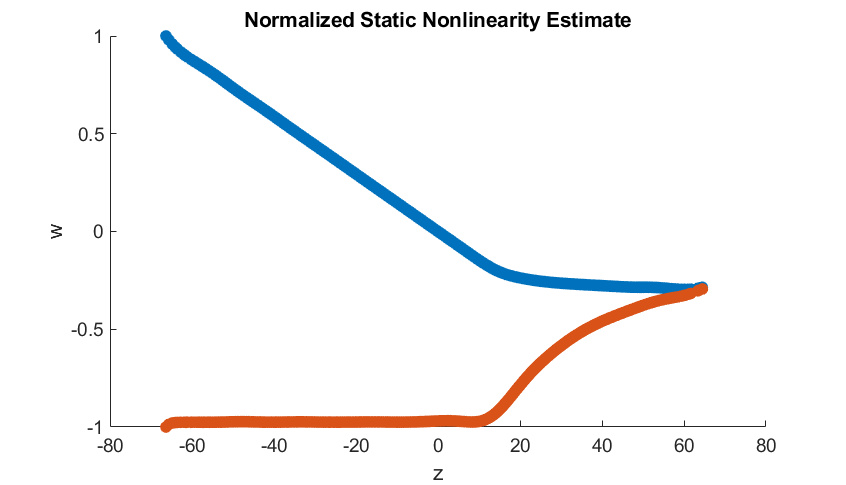}
			\caption{The static nonlinearities (normalized) estimated for the Wiener-Hammerstein model (${f}(z)$, blue) and the scheduling map ($\bar{f}(z)$, red).}
			\label{fig:SNLWH}
		\end{figure}

\begin{figure}[hbt]
			\centering
				\includegraphics[width=0.95\columnwidth]{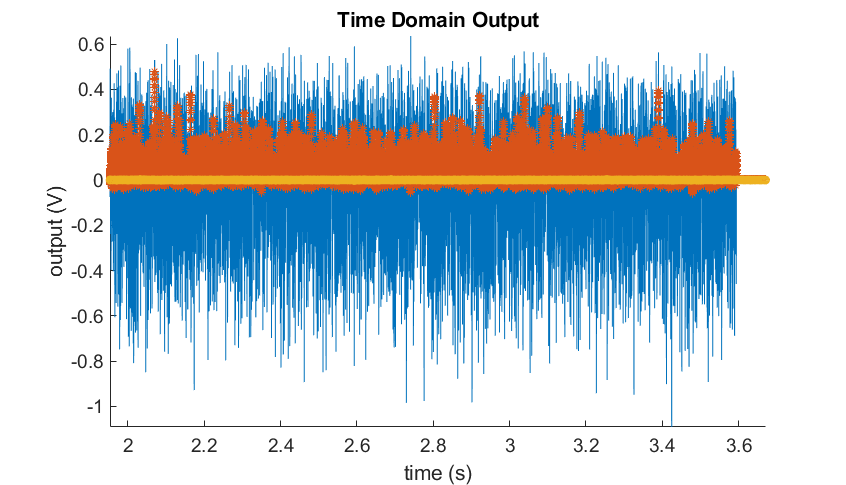}
			\caption{Time domain output of the Wiener-Hammerstein validation data (blue). The error obtained using the LTI approximation (red), and the error of the LPV model (orange).}
			\label{fig:TimeWH}
		\end{figure}

\begin{figure}[hbt]
			\centering
				\includegraphics[width=0.95\columnwidth]{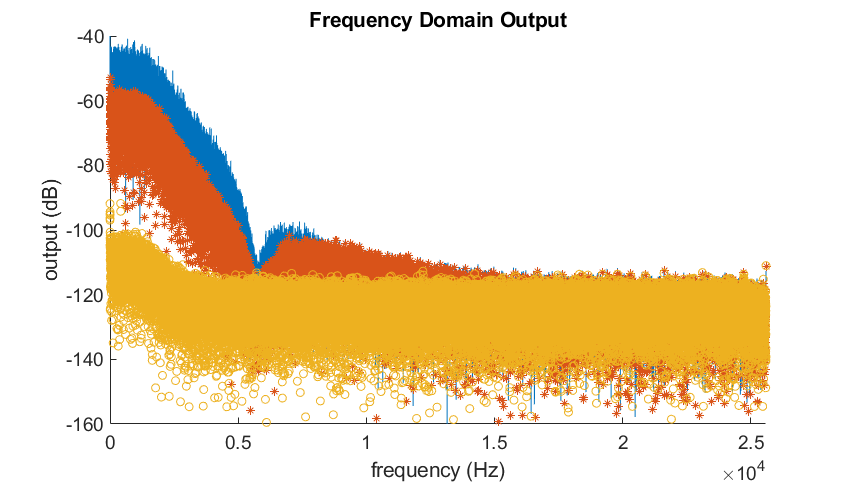}
			\caption{Frequency domain output of the Wiener-Hammerstein validation data (blue). The error obtained using the LTI approximation (red), and the error of the LPV model (orange).}
			\label{fig:FreqWH}
		\end{figure}

		The original identified static nonlinearity ${f}(z(t))$ and the factorized version $\bar{f}(z(t))$ are shown in Figure~\ref{fig:SNLWH}.
		The obtained estimation and validation RMSEs are reported in Table~\ref{tab:RMSEWH}. Note that the obtained LPV representation of the Wiener-Hammerstein benchmark system has a high accuracy, it obtains one of the best RMSE that are reported in the literature \citep{Marconato2013}. A small difference in RMSE can be observed between the nonlinear and LPV model due to the re-estimation of the static nonlinearity (see Section~\ref{sec:WHEmbedding}). A large improvement can be observed by using the embedded LPV model compared to the LTI model, both in the time (Figure~\ref{fig:TimeWH}) and the frequency domain (Figure~\ref{fig:FreqWH}). The error signal of the nonlinear model is not plotted since it almost coincides with the error of the LPV model.
		
		\begin{table}\centering
		\caption{RMSE of the estimated linear (LTI), Wiener-Hammerstein (NL) and LPV model on the estimation and validation dataset. A 1000 point transient is removed to compute the RMSE on the estimation dataset.}
		\begin{tabular}{|c|c|c|c|} 
			 			     &  LTI + offset & NL & LPV \\ 
			\hline 
			 RMSE est. (mV) &  42.2737 & 0.33461 & 0.33258 \\ 	
			 RMSE val. (mV) &  43.2786 & 0.34015 & 0.34088 \\ 
		\end{tabular} 
		 \label{tab:RMSEWH}
		\end{table}
		
\section{Silverbox Benchmark} \label{sec:Silverbox}
	The Silverbox system is of the form \citep{Wigren2013}:
	\begin{align}
		\begin{split}
			y(t) &= G(q)x(t), \\
			x(t) &= u(t)-g(y(t)),
		\end{split}
	\end{align}
	where $g(.)$ is a third order polynomial function. This is very similar to the LFR form presented in Eq.~\eqref{eq:LFRSS1}. Indeed, such a system can be represented by a LFR model where $G^{[i]}(q) = G(q)$ for $i=1,2,3,4$ and $f(.) = g(.)$.
	
	Noise is only present at the output of the system (SNR$\approx$50dB).	
	
	\subsection{Nonlinear Identification}
		First a Silverbox model is identified using the method presented in \cite{Paduart2004,Paduart2008}. This approach first estimates a linear approximation of the nonlinear system and the static nonlinearity is estimated in a second step. The obtained Silverbox model is cast into a LFR form next and optimized further using standard nonlinear least square optimization techniques.
		
		The linear dynamics $G(q)$ are second order dynamics, containing a direct term. To avoid the direct term in the final model, and approximate 4/2 order model is estimated in combination with a 1-sample delay.
		
		A constant output offset $\hat{c}_y$ is also estimated during the nonlinear modeling step.
		
		The following model is estimated:
		\begin{align} \label{eq:ghat}
			\begin{split} 
				\hat{y}(t) &\!=\! \hat{y}_0 + \hat{c}_y, \\
				\hat{y}_0(t) &\!=\! \hat{G}(q)[u(t)-{f}(\hat{y}_0(t))] \\
				\hat{G}(q) &\!=\! \tfrac{0.488 q^{-1} - 0.110 q^{-2} + 0.131 q^{-3} - 0.079 q^{-4} + 0.026 q^{-5}}{0.994 - 1.518 q^{-1} + 0.929 q^{-2}} \\
				{f}(y(t)) &\!=\! 0.0079 \!+\! 0.1166 y(t) \!-\! 0.0060 y^2(t) \!+\! 3.8885 y^3(t) \\
				\hat{c}_y &\!=\! 0.0024.
			\end{split}
		\end{align}
		
		The initial state-space representation of the nonlinear LFR model is of order 20 due to the transformation of the true model structure into the nonlinear LFR model representation (see beginning of Section~\ref{sec:Silverbox}). The model order of the nonlinear LFR model is reduced to 5 using the \texttt{minreal} command in Matlab. 
		
		Similar to the Wiener-Hammerstein case, a number of non-uniqueness transformations are present in the LFR model obtained for the Silverbox system \citep{SchoukensM2017b}. However, the impact of these non-uniqueness transformations are not explored further in this paper.
		
	\subsection{LPV Embedding}
		The resulting LPV model parameters are given by (up to 4 digits):
			\begin{align}
				A &= \left[\begin{array}{cccccc} 
						    0.6483  & -0.7159  & -0.1035  &  0.3406  & -0.7618 \\
						    0.5610  &  0.9673  & -0.3287  &  0.3172  & -0.3435 \\
						    0.0134  & -0.0272  &  0.0462  &  0.0270  & -0.0168 \\
						   -0.0090  &  0.0190  & -0.9964  &  0.0003  &  0.0291 \\
						    0.0055  & -0.0112  &  0.3036  & -1.1280  & -0.1350 \end{array}\right], \nonumber\\
				A_\mathrm{p} &= \left[\begin{array}{cccccc} 
						    0.2296  &  0.1728  & -0.2288  &  0.0997  & -0.2195 \\
						    0.0036  &  0.0027  & -0.0036  &  0.0016  & -0.0035 \\
						   -0.1527  & -0.1149  &  0.1522  & -0.0663  &  0.1460 \\
						    0.1055  &  0.0794  & -0.1052  &  0.0458  & -0.1009 \\
						   -0.0627  & -0.0472  &  0.0625  & -0.0272  &  0.0599 \end{array}\right], \nonumber
			\end{align}
			\begin{align}
			 	B_\mathrm{u} = \left[\begin{array}{c}     0.5219 \\ 0.0082 \\ -0.3471 \\	    0.2398 \\ -0.1425 \end{array}\right], \;			
				B_\mathrm{p} = \left[\begin{array}{c} 0 \\ 0 \\ 0 \\ 0 \\ 0 \end{array}\right], \; D_\mathrm{yu} = 0, \; D_\mathrm{p} = 0, \nonumber
			\end{align}
			\begin{align}
				C_\mathrm{y} &= \left[\begin{array}{cccccc}0.4398  &  0.3311 &  -0.4385   & 0.1911 &  -0.4205 \end{array}\right], \nonumber\\
				C_\mathrm{p} &= \left[\begin{array}{cccccc}0& 0& 0& 0& 0&  0 \end{array}\right], \nonumber
			\end{align}
		
		The scheduling map is given by:
		\begin{align}
			\bar{f}(z(t)) &= 0.1166 - 0.0060 y(t) + 3.8885 y^2(t).
		\end{align}
		The constant offset $0.0079$ that was present in the estimated static nonlinearity (Eq.~\eqref{eq:ghat}) is moved to the input and the output of the model as is explained in Section~\ref{sec:Embedding}.
		
		Note that in this case (see also Section ~\ref{sec:remarks}), the signal $z(t)$ is equal to the nonlinear LFR model output $\tilde{y}(t)$ (Eq.~\eqref{eq:LFRSS4}). 
	
	\subsection{Results}
\begin{figure}[hbt]
			\centering
				\includegraphics[width=0.95\columnwidth]{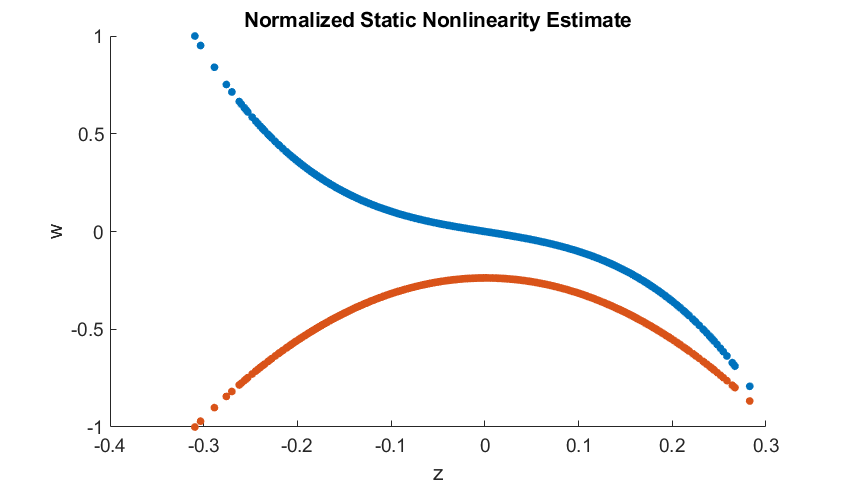}
			\caption{The static nonlinearities (normalized) estimated for the Silverbox model (${f}(z)$, blue) and the scheduling map ($\bar{f}(z)$, red).}
			\label{fig:SNLSB}
		\end{figure}

\begin{figure}[hbt]
			\centering
				\includegraphics[width=0.95\columnwidth]{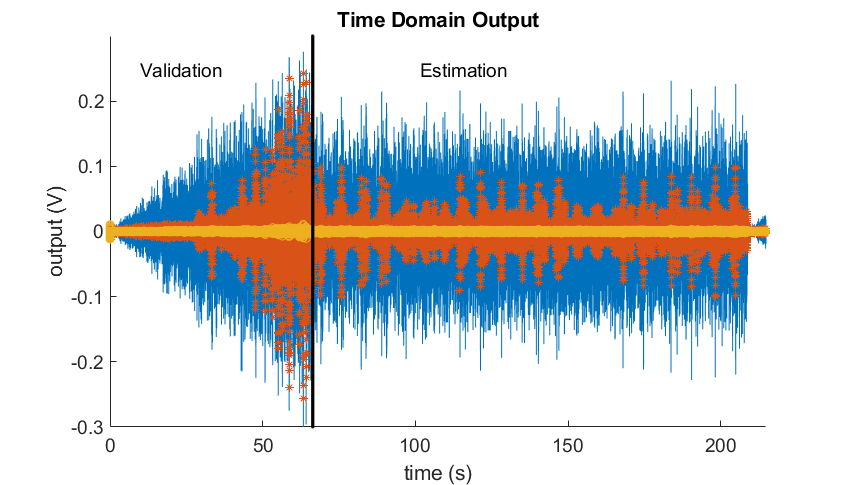}
			\caption{Time domain output of the Silverbox validation and estimation data (blue). The error obtained using the LTI approximation (red), and the error of the LPV model (orange).}
			\label{fig:TimeSB}
		\end{figure}

\begin{figure}[hbt]
			\centering
				\includegraphics[width=0.95\columnwidth]{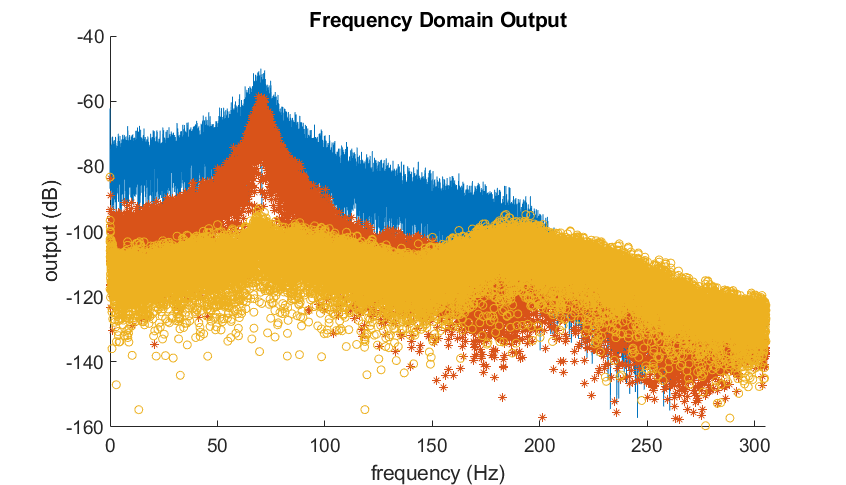}
			\caption{Frequency domain output of the Silverbox validation data (blue). The error obtained using the LTI approximation (red), and the error of the LPV model (orange).}
			\label{fig:FreqSB}
		\end{figure}

		The original identified static nonlinearity ${f}(z(t))$ of the nonlinear model and its factorized version $\bar{f}(z(t))$ are shown in Figure~\ref{fig:SNLSB}.
		The obtained estimation and validation RMSE are reported in Table~\ref{tab:RMSESB}. Note that the obtained LPV representation of the Silverbox benchmark system is of very high quality the results are in line with the ones reported in \cite{Paduart2008}.
		A large improvement can be observed by using the embedded LPV model compared to the LTI model, both in the time (Figure~\ref{fig:TimeSB}) and the frequency domain (Figure~\ref{fig:FreqSB}). The error signal of the nonlinear model is not plotted since it coincides with the error of the LPV model.
		
		\begin{table}\centering
		\caption{RMSE of the estimated linear (LTI), Silverbox (NL) and LPV model on the estimation and validation dataset. A 1000 point transient is removed to compute the RMSE on the estimation dataset.}
		\begin{tabular}{|c|c|c|c|} 
	     &  LTI + offset & NL & LPV \\ 
			\hline 
	 	RMSE est. (mV) &  8.1178 & 0.70646 & 0.70646 \\ 	
	 	RMSE val. (mV) &  16.222 & 0.77604 & 0.77604 \\ 
		\end{tabular} 
		 \label{tab:RMSESB}
		\end{table}
		
\section{Conclusion} \label{sec:Conclusions}
	This paper shows how the class of nonlinear LFR systems can be exactly represented by an affine state-space LPV model without introducing any new singularities. The effectiveness of the proposed approach is illustrated on two established benchmark examples. Constructing the LPV model of a nonlinear system starting from an initial nonlinear model simplifies the selection of the scheduling signal significantly.

	Further research will explore the modeling of nonlinear systems containing multiple static nonlinearities, taking into account the control objectives in the modeling process, and the control of nonlinear systems based on the obtained LPV models.

% bibliography
\bibliography{../../../References/ReferencesLibraryV2}                                

\end{document}